\documentclass[
reprint,
superscriptaddress,
amsmath,amssymb,
aps, pra
]{revtex4-1}

\usepackage{graphicx}
\usepackage{dcolumn}
\usepackage{gensymb}
\usepackage[]{hyperref}
\hypersetup{colorlinks,
	linkcolor={blue!75!black!80!yellow},
	citecolor={blue!75!black!80!yellow},
	urlcolor={blue!75!black!80!yellow},
	pdfstartview=FitH}
\usepackage{xcolor}
\usepackage{enumitem}
\usepackage{upgreek}

\usepackage{color,soul} 

\usepackage[resetlabels,labeled]{multibib}

\begin{document}

\title{Resolving photon numbers using a superconducting tapered nanowire detector}

\author{Di Zhu}\thanks{These authors contributed equally}

\author{Marco Colangelo}\thanks{These authors contributed equally}

\author{Changchen Chen}

\affiliation{Research Laboratory of Electronics, Massachusetts Institute of Technology, Cambridge, Massachusetts 02139, USA}

\author{Boris A. Korzh}\affiliation{Jet Propulsion Laboratory, California Institute of Technology, Pasadena, California 91109, USA}

\author{Franco N. C. Wong}\affiliation{Research Laboratory of Electronics, Massachusetts Institute of Technology, Cambridge, Massachusetts 02139, USA}

\author{Matthew D. Shaw}
\affiliation{Jet Propulsion Laboratory, California Institute of Technology, Pasadena, California 91109, USA}

\author{Karl K. Berggren}\email{berggren@mit.edu}
\affiliation{Research Laboratory of Electronics, Massachusetts Institute of Technology, Cambridge, Massachusetts 02139, USA}

\date{November 20, 2019}

\begin{abstract}
Time- and number-resolved photon detection is crucial for photonic quantum information processing. Existing photon-number-resolving (PNR) detectors usually have limited timing and dark-count performance or require complex fabrication and operation. Here we demonstrate a PNR detector at telecommunication wavelengths based on a single superconducting nanowire with an integrated impedance-matching taper. The prototyping device was able to resolve up to five absorbed photons and had 16.1\,ps timing jitter, $<$2\,c.p.s. device dark count rate, $\sim$86\,ns reset time, and 5.6\% system detection efficiency (without cavity) at 1550 nm. Its exceptional distinction between single- and two-photon responses is ideal for coincidence counting and allowed us to directly observe bunching of photon pairs from a single output port of a Hong-Ou-Mandel interferometer. This detector architecture may provide a practical solution to applications that require high timing resolution and few-photon discrimination.
\end{abstract}


\maketitle
The ability to resolve the photon number of an optical field with precise timing is desirable for many applications in quantum information science, including linear optical quantum computing~\cite{Kok2007}, quantum key distribution~\cite{Brassard2000, Lo2005}, quantum repeaters~\cite{Krovi2016}, and non-classical state generation~\cite{Waks2006}. Significant effort has been made to develop photon-number-resolving (PNR) detectors~\cite{Mirin2012,Eisaman2011}, but their performance, especially at telecommunication wavelengths, is often limited in terms of timing resolution~\cite{Lita2008}, reset time~\cite{Gansen2007}, dark count rate~\cite{Jiang2007}, and PNR fidelity~\cite{Kardyna2008}. 

Superconducting nanowire single-photon detectors (SNSPDs) are currently the leading single-photon counting technology at near-infrared wavelengths, with $>90$\% efficiency, sub-3-ps jitter, few-ns reset time, and sub-Hz dark count rate~\cite{Hadfield2009,Natarajan2012,Holzman2019}. However, unlike transition-edge sensors (TES)~\cite{Lita2008} or microwave kinetic inductance detectors (MKID)~\cite{Day2003}, SNSPDs operate in a highly nonlinear mode and lack intrinsic photon number resolution. To overcome this problem, past efforts mainly focused on implementing arrays that consist of multiple closely-packed nanowires, each detecting one photon~\cite{Dauler2009b, Divochiy2008}.  They can be read out through certain multiplexing schemes but usually require complex fabrication~\cite{Divochiy2008, Marsili2009,Jahanmirinejad2012a,Cheng2012} or signal processing~\cite{Zhu2018}. Moreover, to avoid multiple photons hitting the same element, the array size needs to be much larger than the input photon number~\cite{Jonsson2019, Dauler2009b}. These architectural limits have hindered the use of SNSPD arrays in applications requiring photon number resolution.

Closer scrutiny of the detection mechanism suggests that the lack of PNR capability in SNSPDs may not be intrinsic. In 2007, Bell et al. recognized that $n$-photon absorption in a long meandered superconducting nanowire should induce $n$ resistive hotspots ($n$ is an integer)~\cite{Bell2007}. However, the resistance change due to different numbers of hotspots is hardly observable because of the abrupt mismatch between the k$\Omega$ resistance of the hotspots and the 50 $\Omega$ impedance of the readout circuit. More specifically, regardless of $n$, the 50 $\Omega$ load will always divert most of the bias current in the nanowire, since $n\,\mathrm{k}\Omega/(n\,\mathrm{k}\Omega + 50\,\Omega)\approx 1$; and therefore, the output voltage remains almost constant.  While it is possible to develop a high-impedance cryogenic readout to avoid this limitation~\cite{Kitaygorsky2009,Cahall2018}, the load impedance must be kept low to prevent latching effects~\cite{Annunziata2010, Kerman2013}; otherwise, active resets will be needed. As a result, matching  the readout to the hotspot resistance remains impractical. Alternatively, Cahall et al. studied the rising edge slope of the detector pulses instead of the output amplitude and observed faster slew rates for multi-photon events~\cite{Cahall2017}. This method demonstrated the PNR capability in a conventional SNSPD, but the resolution was largely limited by the signal-to-noise ratio and variations of hotspot resistances.

Recently, we developed an impedance-matching technique for SNSPDs based on tapered transmission lines~\cite{Zhu2019}. The taper can provide the SNSPD with a k$\Omega$ load impedance without latching while interfacing the readout electronics at 50 $\Omega$. Here, we use it to make the SNSPD output amplitude sensitive to the number of photon-induced hotspots, and thus enable more practical photon number resolution. This architecture does not require multi-layer fabrication or complex readout, and offers significant advantages over array-type PNR detectors. Though the output amplitude scales sub-linearly with photon numbers, the distinction between single- and multi-photons is exceptionally large. Such a ``few-photon'' detector is especially important for heralding single-photon sources and improving the security of quantum cryptography~\cite{Brassard2000}.

\begin{figure*}
    \centering
    \includegraphics[]{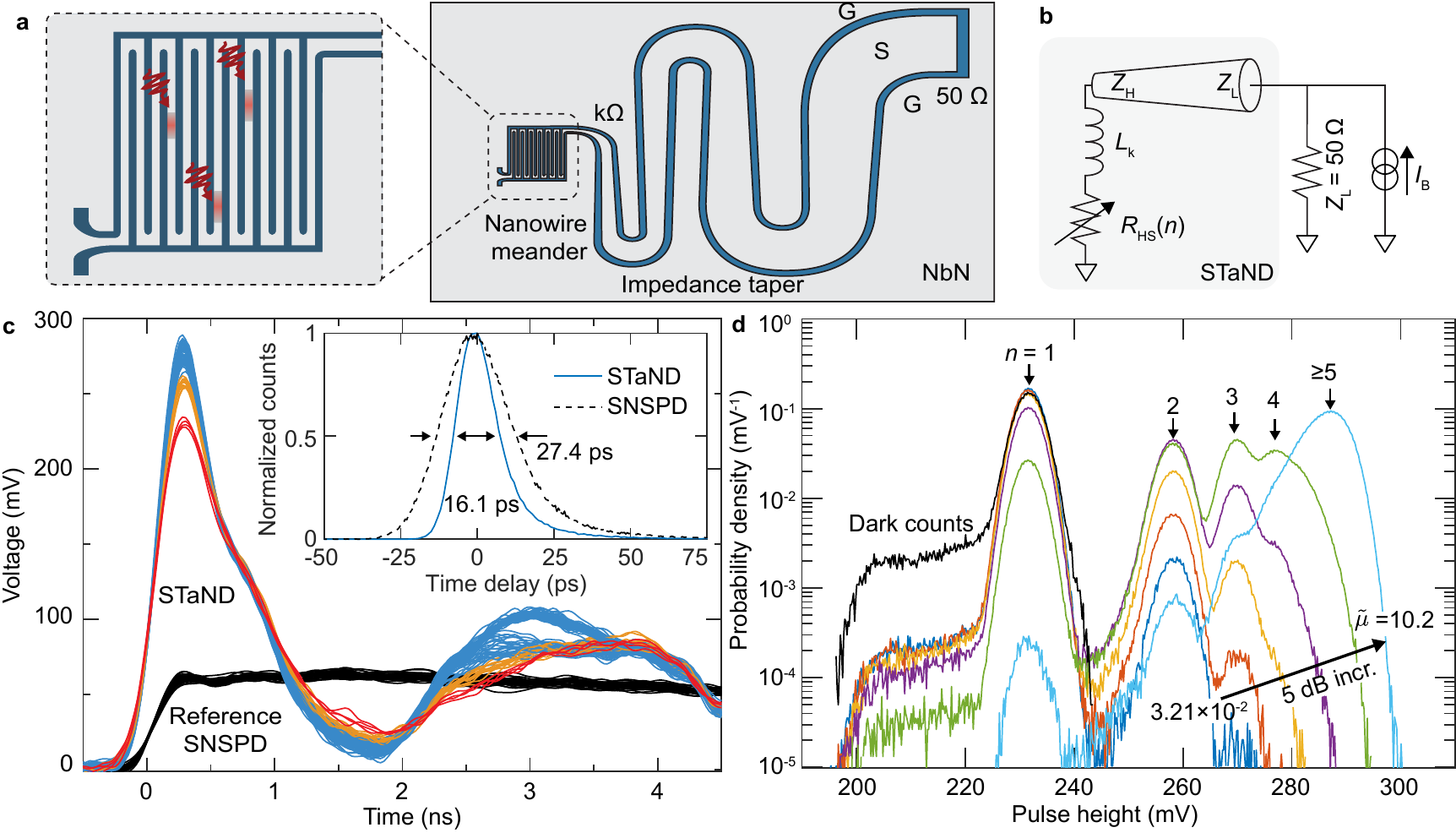}
    \caption{\textbf{Superconducting tapered nanowire detector (STaND).} \textbf{a}, The STaND consists of a photon-sensitive nanowire meander and a transmission line taper (drawing not to scale), whose characteristic impedance transitions from k$\Omega$ to 50 $\Omega$. Zoomed panel: $n$-photon absorption induces $n$ hotspots in the nanowire meander. Grey: superconducting NbN thin film; blue: substrate (SiO$_2$ on Si), where NbN was etched away. \textbf{b}, Equivalent circuit diagram of STaND. The variable resistor $R_\mathrm{HS}(n)$ captures the photon-number-dependent hotspot resistance (k$\Omega$ scale). The taper provides an effective load impedance ($Z_\mathrm{H}$) that is comparable to $R_\mathrm{HS}(n)$, making the output voltage sensitive to $n$. \textbf{c}, Compared to a regular SNSPD (black traces), the output pulses from the STaND (colored, consisting of 80 traces acquired at $\tilde{\mu}=4.04$) not only have larger amplitudes but also showed level separation for multi-photon events. Red: single-photon events; orange: two-photon events; blue: three- or higher-photon events. Inset: instrument response functions at 1550 nm in the single-photon regime, where STaND showed a FWHM timing jitter of 16.1 ps as compared to 27.4 ps for the reference SNSPD.  \textbf{d}, Histograms of the STaND's pulse heights under pulsed laser illumination. Each photon counting histogram was constructed from $10^6$ detection events, while the dark count histogram (black curve) was from $2\times10^5$ events.}
    \label{fig:fig1}
\end{figure*}

The basic architecture and key features of the superconducting tapered nanowire detector (STaND) are summarized in Fig.~\ref{fig:fig1}. The STaND consists of two parts: (1) a photon-sensitive nanowire meander (similar to a conventional SNSPD), and (2) an impedance-matching taper, whose characteristic impedance gradually decreases from $Z_\mathrm{H}\approx 2.4\, \mathrm{k}\Omega$ on the narrow end to $Z_\mathrm{L}=50\,\Omega$ on the wide output end (Fig.~\ref{fig:fig1}a, drawing not to scale; see Supplementary Fig.~S1 for micrographs of the device). The STaND can be represented using an equivalent circuit shown in Fig.~\ref{fig:fig1}b. The nanowire meander is modeled as a photon-number-dependent variable resistor $R_\mathrm{HS}(n)$ in series with a kinetic inductor $L_\mathrm{K}$~\cite{Kerman2006a}, and the taper is modeled as a multi-section impedance transformer. $n$ absorbed photons will induce $n$ initial hotspots, which then expand through electrothermal feedback in the nanowire~\cite{Kerman2009} (Fig.~\ref{fig:fig1}a inset). In general, $R_\mathrm{HS} (n)$ is on the order of k$\Omega$ and increases with $n$ (but scales sub-linearly, see Supplementary Fig.~S4). Following a simple division rule, the current leaving the nanowire (entering the taper) scales roughly as $\sim R_\mathrm{HS}(n)/(R_\mathrm{HS}(n)+Z_\mathrm{H})$. Note that this dependence on $n$ becomes appreciable only when $Z_\mathrm{H}$ is comparable to $R_\mathrm{HS}$, a condition that could not be achieved without the taper. Moreover, the taper acts as a transformer that increases the current into the load, making the output voltage larger by as much as $\sqrt{Z_\mathrm{H}/Z_\mathrm{L}}$ (in the case of perfect power transfer) compared to a 50 $\Omega$-loaded standard SNSPD ~\cite{Zhu2019}. The simplistic picture described here ignores the complex electrothermal feedback and microwave dynamics in the nanowire and taper~\cite{Zhu2019, Zhao2018a}. In the Supplementary Information (SI), we simulate the hotspot evolution and pulse shapes using a SPICE model that incorporates both effects~\cite{Zhu2019, Zhao2018a, Berggren2018}.

Figure~\ref{fig:fig1}c compares the output waveforms from the STaND and a reference SNSPD with the same nanowire meander design on the same chip. The STaND output not only had larger amplitudes ($>3.6$ times) and faster slew rates ($>4$ times)~\cite{Zhu2019}, but also exhibited level separations from multi-photon events (colored according to maximum pulse height for clear visualization). The faster slew rate reduced timing jitter from 27.4 ps full-width at half-maximum (FWHM) to 16.1 ps at 1550 nm (see Fig.~\ref{fig:fig1}c inset). Other than amplitude difference, the pulse shapes were also observed to exhibit distinct signatures due to microwave reflections in the nanowire and taper. For example, the blue traces were separated at $t=3$ ns, making three- and higher-photon events distinguishable. We processed the pulses' rise times and slew rates and found correlations to photon numbers as well, which agreed with recent reports on the photon-number-dependent slew rates~\cite{Cahall2017} (see Supplementary Fig.~S7).

We probed the multi-photon response of the STaND using an attenuated 1550 nm pulsed laser. Figure~\ref{fig:fig1}d shows histograms of the pulse heights at effective mean photons per pulse $\tilde{\mu}$ ranging from $3.21\times10^{-2}$ to $1.02\times10^1$. Here, $\tilde{\mu}=\eta \mu$, where $\mu$ is the mean photon per pulse of the coherent source, and $\eta$ includes the coupling losses and detector efficiencies (see the Methods section for details on the estimation of $\tilde{\mu}$). We observed level separations up to five photons. When $n\ge5$, the levels were no longer separable, and further increasing $\tilde{\mu}$ only gradually shifted the peak position. The shoulders in the histograms (pulse heights $<$ 220 mV) were likely from counting events at the bends in the nanowire meander. The dark count histogram (black line) showed a more prominent shoulder because current tended to crowd in the bends and generate more dark counts (see Supplementary Fig.~S9). We further confirmed this hypothesis by moving the fiber focuser to illuminate more on the bends and corners, and observed more prominent shoulders (see Supplementary Fig.~S8). 

Figure~\ref{fig:fig2}a and b show Gaussian fits of pulse height histograms at $\tilde{\mu}$ = 1.01 and 3.19. The separation between single- and two-photon peaks (26.4 mV) was more than 10.7 standard deviations of their spread (10.7$\sigma$), making the STaND suitable for single-shot discrimination between single- and multi-photons. The widths of the Gaussians (5.5 mV FWHM for $n=1$ and $>5.8$ mV FWHM for $n\ge2$) were larger than the measured electrical noise floor in the system (4.2 mV FWHM, see Supplementary Fig.~S6), suggesting the existence of other fluctuation mechanisms or inhomogeneities, such as variation in nanowire width, hotspot location, and photon inter-arrival time. In addition, the pulse peaks could be partially low-pass filtered and under-sampled due to the limited bandwidth of the amplifier chain and the finite resolution of the oscilloscope.

\begin{figure}
    \centering
    \includegraphics[width = 3.4in]{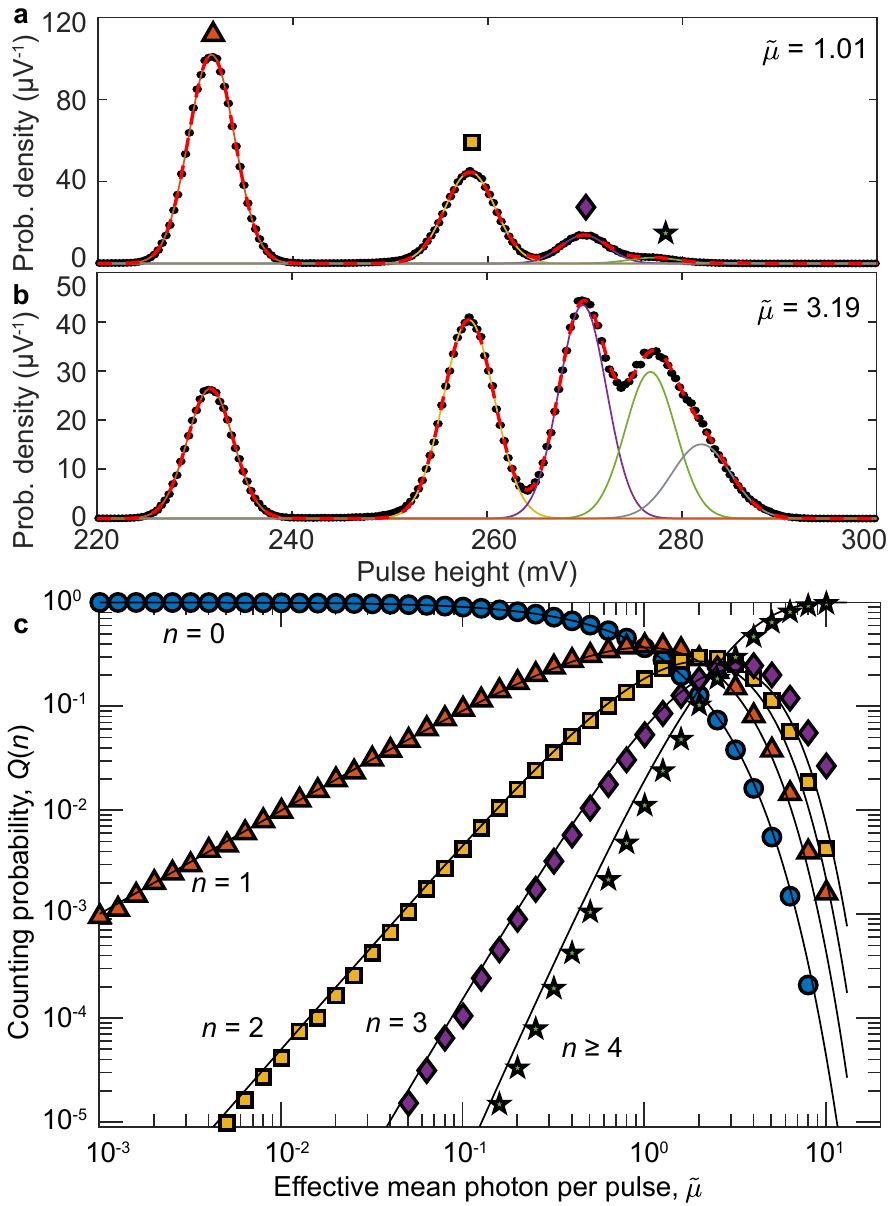}
    \caption{\textbf{Counting statistics under coherent state illumination.} \textbf{a, b}, Gaussian fitting of the pulse height histograms when the STaND was illuminated using a pulsed laser with $\tilde{\mu}$ at 1.01 and 3.19, respectively. Black dots: measurement data; dashed red line: fitting result; solid lines: decomposed Gaussian functions.  \textbf{c}, Photon counting statistics reconstructed from the pulse height distributions. $Q(n)$ is the probability of detecting an $n$-photon event. The measured counting statistics (symbols) directly followed the Poisson statistics of the coherent source, $S(n) = e^{-\tilde{\mu}} \tilde{\mu}^n/n!$ (lines).}
    \label{fig:fig2}
\end{figure}

We integrated the area under each Gaussian curve to reconstruct the counting statistics, $Q(n)$ (as shown in Fig.~\ref{fig:fig2}c). Here, we  grouped $Q(n\ge 4)$ as these events were not well-separated. In the STaND, since the total length of the nanowire ($\sim 
 500\,\upmu\mathrm{m}$) was about 1000 times longer than the hotspot size (on the order of 100 nm), the probability of hotspot overlap was negligible.  Therefore, in general, $Q(n)$ directly followed the Poisson statistics of the laser source, $S(n) = e^{-\tilde{\mu}}\tilde{\mu}^n/n!$ without the need of a conditional probability, which is required in most array-type PNR detectors to account for the possibility of multiple photons hitting the same element~\cite{Dauler2009b,Divochiy2008}.  
 
However, the photons need to overlap in time. More specifically, when the first photon is absorbed, the current in the nanowire starts to drop rapidly (80\% to 20\% time constant of $\sim 200$ ps, inferred from the detector rise time, see Supplementary Fig.~S7); if the second photon arrives late, it will be detected with a lower probability due to the decreased bias current. This dependence on inter-arrival time is likely why $Q(n)$ in Fig.~\ref{fig:fig2}c tends to be under-estimated when $n$ is large (the laser pulse used here had a FWHM of 33 ps). When we further increased the laser pulse width, the under-estimation became more prominent (see Supplementary Fig.~S11). Therefore, the STaND can only resolve multiple photons within a tens of ps time window, which is acceptable and sometimes desirable for many applications.

\begin{figure*}
    \centering
    \includegraphics[]{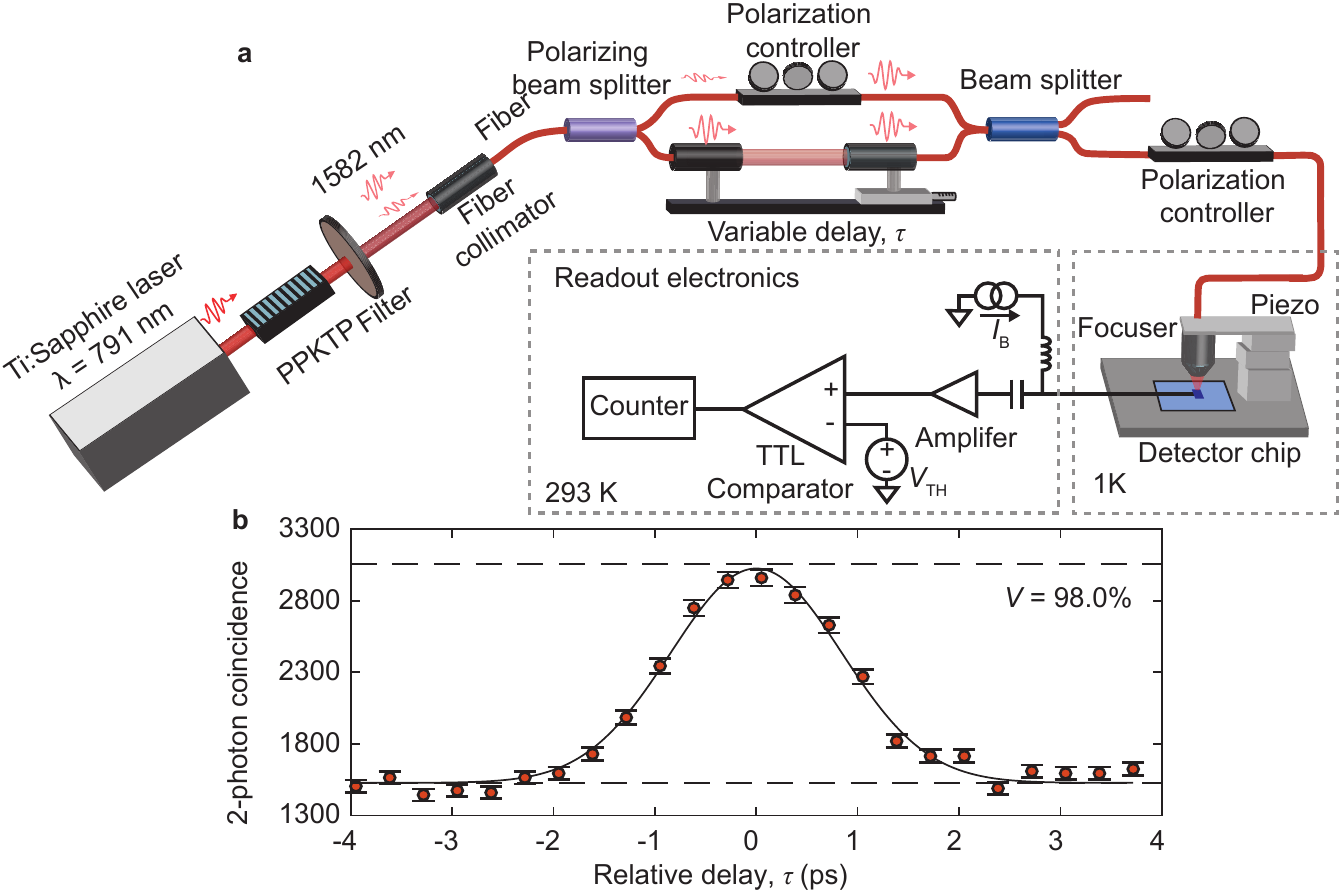}
    \caption{\textbf{Measuring Hong-Ou-Mandel (HOM) interference using a single STaND.} \textbf{a}, Experimental setup. Frequency-entangled photon pairs were generated in a type-II phase-matched PPKTP crystal and separated based on their polarization. They then interfered at a 50:50 beam splitter with a relative time delay $\tau$, and the coincidence was monitored using a STaND at a single output port instead of two single-photon detectors at both output ports. The amplified detector pulses went through a comparator, whose threshold voltage $V_\mathrm{TH}$ was controlled by a programmable battery source and set to only register multi-photon events. \textbf{b}, Coincidence counts (background subtracted) in a 300 s time window as a function of relative delay between the two photons, showing clear bunching with interference visibility of 98\% (well above the 50\% classical limit). Error bar: one standard deviation due to Poisson noise; solid curve: Gaussian fit; dashed lines: baseline and theoretical ceiling for perfect interference (twice the baseline).}
    \label{fig:fig3}
\end{figure*}

Next, we present the STaND's direct application in measuring non-classical states of light. When two indistinguishable photons interfere at a beam splitter, they tend to leave from the same output port (bunching), a phenomenon known as Hong-Ou-Mandel (HOM) interference~\cite{Hong1987}. This effect is usually demonstrated using two single-photon detectors, one at each output port, and a coincidence dip between the two implies that both photons leave from the same port. Here, we use the STaND to directly observe photon bunching in HOM interference from a single output port of the beam splitter~\cite{DiGiuseppe2003}.

Figure~\ref{fig:fig3}a shows the experimental setup for the HOM interference. Frequency-entangled photon pairs were generated through spontaneous parametric down-conversion (SPDC) process and were separated using a polarizing beam splitter (see Methods section for details)~\cite{Chen2017, Chen2019}. The signal photon went through a tunable air-gap delay ($\tau$), and the idler photon polarization was rotated by 90$^\circ$. The two photons then interfered at a 50:50 beam splitter, and their coincidence as a function of $\tau$ was monitored at a single output port using a STaND. We used a comparator for level discrimination, which generated a pulse only when the input exceeded the threshold voltage $V_\mathrm{TH}$. 

Figure~\ref{fig:fig3}b shows the coincidence counts (300s integration time) as a function of the relative time delay $\tau$. It shows clear photon bunching with interference visibility $V$ of $98.0\pm 3.0\%$ (uncertainty indicates 95\% confidence bound of the Gaussian fitting). $V$ is defined as $(N_\mathrm{max} - N_\mathrm{min})/N_\mathrm{min}$, where $N_\mathrm{max/min}$ represents the maximum (minimum) coincidence counts. Here, we subtracted background two-photon counts ($55\pm7$ from the air-gap delay path and $256\pm16$ from the polarization controller path) measured by blocking individual beam path of the interferometer. These background two-photon counts were caused by imperfections in polarization controls and multi-pair events from SPDC process due to high pump power. Without subtraction, the raw coincidence had visibility of $81.4\pm2.8\%$. In the experiment, we did not observe any two-photon dark counts (pump laser blocked) during the integration time, which ensured a high signal-to-noise ratio for the measurement. The fact that high visibility HOM interference can be observed with a single detector validates the effectiveness of the photon number resolution of the STaND in practical quantum photonic applications, and such measurement can be used to characterize the indistinguishability of single-photon sources.

The key detector metrics of the STaND demonstrated here include 16.1 ps FWHM timing jitter, 1.7 c.p.s. device dark count rate (26.8 c.p.s. system dark count rate, see Supplementary Fig.~S13), 85.8 ns reset time (estimated from the exponential decay time, 3$\tau$, of the output pulses, see Supplementary Fig.~S14), and  $\sim$5.6\% system detection efficiency at 1550 nm (see Supplementary Fig.~S13), all measured at 1.0 K. The system efficiency is currently limited by optical coupling and can in-principle reach $>$90\% using the widely adopted methods of cavity integration and self-aligned fiber packaging~\cite{Miller2011, Marsili2013}. The reset time is limited by the taper inductance, which can be significantly reduced by using microstrip~\cite{Zhu2018} or grounded coplanar waveguide designs (see Supplementary Fig.~S2). The PNR fidelity and dynamic range may be further improved by using low-noise cryogenic amplifiers and comparators and tapers with higher input impedance. Currently, with $>10\sigma$ separation between single- and multi-photon responses, the STaND is well suited as a coincidence counter. When using arrays of detectors to resolve large numbers of incident photons, an array made of such ``two-photon'' detectors would require $\sim$10 times fewer elements than a click/no-click detector array (assuming unity-efficiency, see Supplementary Fig.~S16).  

In conclusion, we have demonstrated a new detector architecture, STaND, whose output amplitude directly encodes photon numbers. It does not require complex fabrication or readout, and inherits the outstanding detector metrics of existing high-performance SNSPDs. With our ongoing efforts to optimize system efficiency and taper designs as well as to incorporate low-noise cryogenic readouts, we expect the STaND to become a readily accessible technology and find immediate applications, such as heralding or rejecting multi-pair generation in SPDC, characterizing single-photon emitters, and preparing and verifying non-classical states of light.

\section*{Methods}
\paragraph{Device design and fabrication.}
The device design and fabrication followed our previous work~\cite{Zhu2019}, but the taper design was modified to have a higher input impedance (2.4 k$\Omega$) and cut-off frequency (290 MHz) for larger PNR dynamic range and and faster reset. The nanowire meander was 100 nm wide and spanned an area of $11\upmu\mathrm{m}\times10\,\upmu\mathrm{m}$ with 50\% fill factor. The taper was a coplanar waveguide whose center conductor width increased from 300 nm (2.4 k$\Omega$) to 160 $\upmu$m (a fixed gap size of 3 $\upmu$m), following the Klopfenstein profile~\cite{Klopfenstein1956}. The STaND reported in this work had a switching current of 25 $\upmu$A at 1.0 K. The reference SNSPD compared in Fig.~\ref{fig:fig1}c and d were fabricated on the same detector chip with same meander design as the STaND, and had a switching current of 27 $\upmu$A. Measurement data presented in this manuscript were taken from a single set of STaND and reference detector. Similar results were reproduced on a different detector chip from a separate fabrication run, including the photon-number-dependent pulse amplitudes, the increase of output voltage, and the reduction of timing jitter.

\paragraph{Detector measurement.}
The detectors were measured in a closed-cycle cryostat at 1.0 K. Light was coupled to the detector using a fiber focuser ($1/e^2$ diameter $<10\upmu$m), mounted on a piezo-positioner. The detector output was amplified using room-temperature amplifiers only--an LNA2500 followed by an LNA2000 (RF Bay). Because the output of the STaND was too large and saturated the second amplifier, a 16 dB attenuator was added. The amplified detector pulses were then either captured using a 6 GHz oscilloscope (Lecroy 760Zi, 40 G samples/s sampling rate) or a universal counter (Agilent 53132A). In the HOM interference measurement, since the counter had a limited trigger level resolution (5 mV), we used a TTL comparator (PRL-350 TTL, bandwidth: 300 MHz) for level discrimination, and used a programmable battery source (SRS SIM928, 1 mV resolution) to supply the threshold voltage. A 1550 nm modulated pulsed diode laser was used (PicoQuant LDH-P-C-1550 laser head with PDL 800-B driver) to probe the multi-photon response of the STaND. The modulated laser diode was triggered at 100 kHz to avoid capacitive charging on the readout amplifier, which tends to artificially shift the pulse height. For jitter measurement shown in Fig.~\ref{fig:fig1}d, a 1550 nm sub-picosecond fiber-coupled mode-locked laser (Calmar FPL-02CCF) was used. 

\paragraph{HOM interference setup.} 
Frequency-entangled photon pairs were generated from SPDC process in a type-II phase-matched periodically-poled KTiOPO$_4$ (PPKTP) crystal pumped by an 80-MHz mode-locked, $\sim$100 fs (FWHM 7.8 nm) Ti:sapphire laser centered at 791 nm at 90 mW. The crystal was temperature stabilized at 21.4$^\circ$C to yield frequency-degenerate signal and idler output at 1582 nm. After the pump was filtered out by a long pass filter (Semrock BLP02-1319R-25), the orthogonally-polarized signal and idler photons were coupled into a polarization-maintaining (PM) fiber. The signal and idler were separated by a polarization beam splitter and recombined on a 50:50 beam splitter. The polarization of signal and idler photons were made to be the same, and their relative delay on the beam splitter was controlled using a tunable air gap on a translational stage ($\sim$3 dB loss). One output of the beam splitter was connected to the STaND located in a different building while the other output was left unconnected. The polarization of the output photons was tuned for the maximum detector efficiency. 

\paragraph{Estimation of effective mean photon number (per pulse).} 
The STaND can be treated as a spatially-multiplexed, $N$-element ($N$ on the order of 1000), uniform-efficiency ($\eta$) detector array. In the SI, we show that the counting statistics of illuminating a $\eta$-efficiency detector array with $\mu$-mean-photon coherent source is identical to that of illuminating a unity-efficiency detector array with $\eta\mu$-mean-photon coherent source. To estimate the effective mean photon $\tilde{\mu}=\eta\mu$ of the pulsed laser, we swept the optical attenuation ($\gamma$) through a calibrated variable attenuator and measured the photon count rate ($PCR$). By fitting the detection probability (ratio between $PCR$ and laser repetition rate $f_\mathrm{rep}$) as $1-\exp(-\gamma\tilde{\mu})$, we extracted $\tilde{\mu}$. This method automatically captures all losses in the measurement setup without the need for optical power measurement. Using $\tilde{\mu}$ in analyzing results from coherent state illumination (Fig.~\ref{fig:fig1}e and Fig.~\ref{fig:fig2}) helps us isolate the detector's architectural limit on PNR quality (such as sensitivity on photon inter-arrival time and probability of overlapping hotspots) from the unoptimized system efficiency. 

\section*{Acknowledgment}
We thank Qing-Yuan Zhao, Emily Toomey, Brenden Butters, Ilya Charaev, Marco Turchetti, and Neil Sinclair for helpful discussion. We thank the Optical Communications Technology Group at MIT Lincoln Laboratory for their support on the Spectra Physics MaiTai laser. Part of this work was performed at the Jet Propulsion Laboratory, California Institute of Technology, under contract with the National Aeronautics and Space Administration. Support for this work was provided in part by the JPL Strategic University Research Partnerships program, DARPA Defense Sciences Office through the DETECT program, and National Science Foundation grants under contract No. ECCS-1509486. D. Z. was supported by the National Science Scholarship from A*STAR, Singapore.

\end{document}


\widetext
\begin{center}
	\textbf{\large Supplementary Information \\ Resolving photon numbers using a superconducting tapered nanowire detector}
\end{center}
\setcounter{equation}{0}
\setcounter{figure}{0}
\setcounter{table}{0}
\setcounter{page}{1}
\makeatletter
\renewcommand{\theequation}{S\arabic{equation}}
\renewcommand{\thefigure}{S\arabic{figure}}
\renewcommand{\bibnumfmt}[1]{[S#1]}
\renewcommand{\citenumfont}[1]{S#1}

%
%
%
%
%
%
%
%
%
%
%
%


\section{Device layout}
\subsection{Micrographs of the fabricated device}

\begin{figure}[H]
    \centering
    \includegraphics[width=6in]{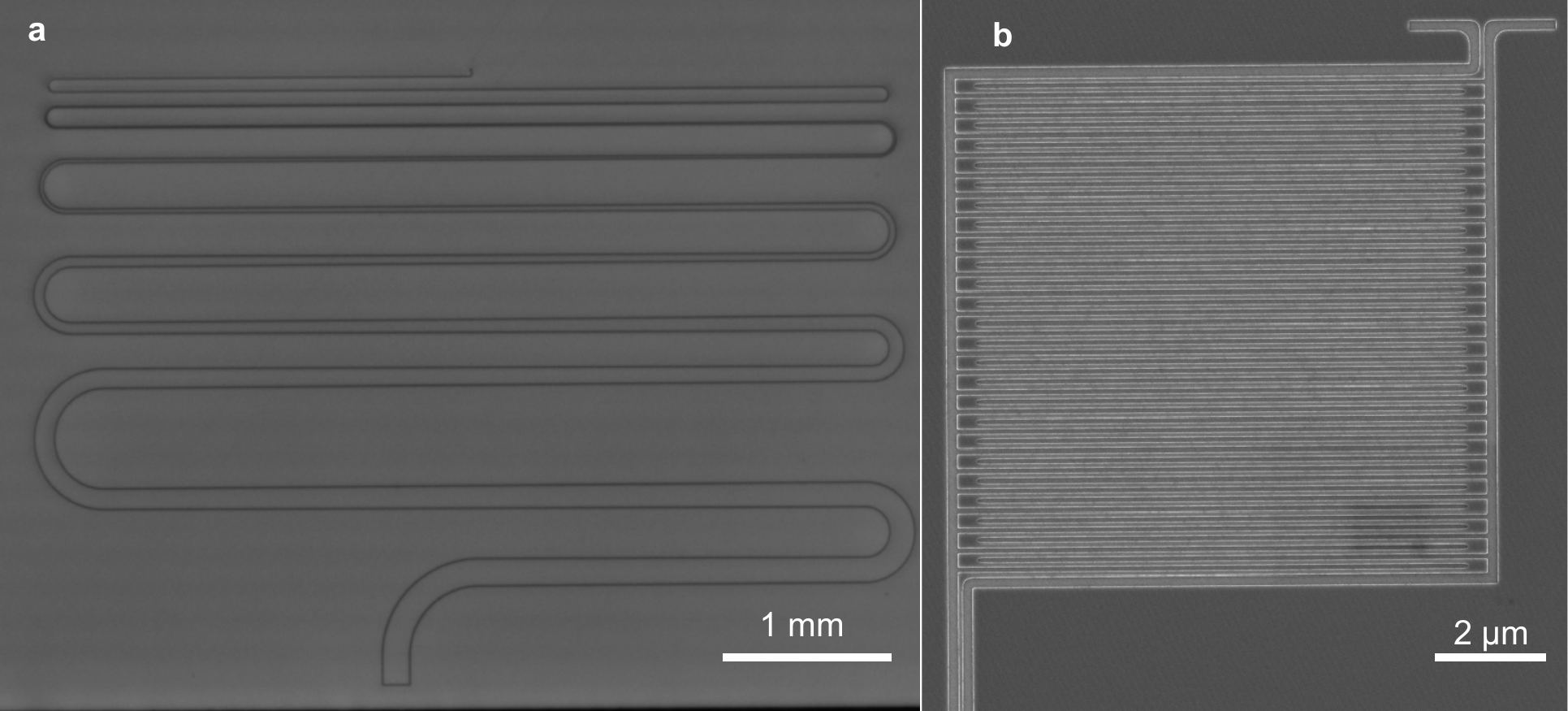}
    \caption{Device layout. (a) Optical micrograph of the taper. Its center conductor starts with 300 nm, and gradually increases to 160 $\upmu$m (50 $\Omega$). The dark outlines are the gap (3 $\upmu$m). The taper size can be significantly reduced by using microstrip or grounded CPW designs. (b) Scanning electron micrograph of the naonwire meander. Dark regions are NbN; bright regions are the substrate, where NbN was etched away.}
    \label{fig:micrograph}
\end{figure}

Figure~\ref{fig:micrograph} shows micrographs of the device presented in the main text. Figure~\ref{fig:micrograph}a shows the over device, where the taper occupies most of the footprint, while Figure~\ref{fig:micrograph}b shows the detail of the nanowire meander.

\subsection{Reducing taper size with grounded CPW design}
\begin{figure}[H]
    \centering
    \includegraphics[]{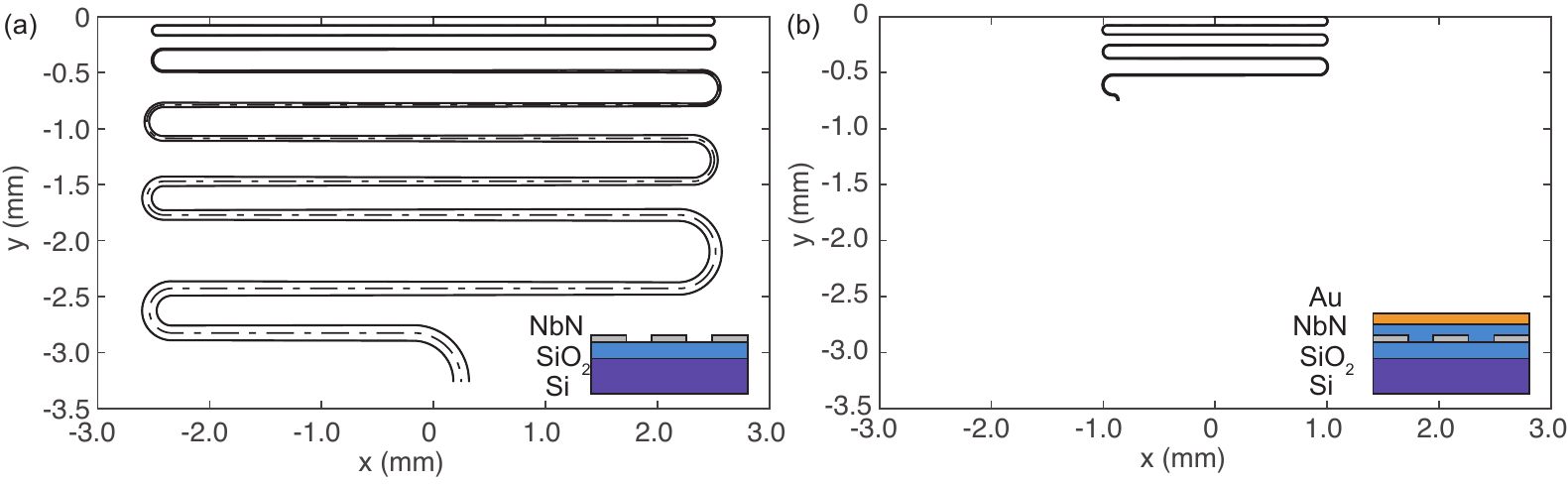}
    \caption{Reducing taper inductance and footprint by using CPW with a top ground. (a) Profile of the CPW taper used in the measured device. (b) Profile of a CPW taper (1 $\upmu$m gap) with top ground (grounded CPW). Adding a gold ground on top of the NbN with a 120-nm-thick SiO$_2$ spacer increases the line inductance and shrinks the size of the taper. Both tapers have the same cut-off frequency (290 MHz) and initial center conductor width (300 nm). The CPW in (a) is 52 mm long and has 16600 squares (1,328 nH assuming 80 pH/sq for the NbN film), while the grounded CPW in (b) is only 11 mm long and has 9118 squares (729 nH). The reduced inductance will shorten the reset time of the detector.}
    \label{fig:cpw_vs_gnd_cpw}

\end{figure}

The taper footprint and inductance can be readily reduced by using microstrip or CPWs with closely placed top or bottom ground (grounded CPW). Figure~\ref{fig:cpw_vs_gnd_cpw} compares the sizes of (a) the CPW taper used in this work and (b) a grounded CPW taper. The calculated grounded CPW taper shown in Fig.~\ref{fig:cpw_vs_gnd_cpw}(b) has a gap size of 1 $\upmu$m, and a top gold ground separated by a 120 nm SiO$_2$ spacer. The substrate is Si with 300 nm thermal oxide. Adding the top ground increases the capacitance per unit length of the transmission line, which increases the effective index and reduces taper size. Both tapers follow Klopfenstein profile~\cite{Klopfenstein1956} and have the same cut-off frequency at 290 MHz. They both start with 300 nm center conductor width, and end with 50 $\Omega$ impedance. The CPW taper has a total length of 52 mm (16,600 squares). Assuming a sheet inductance of 80 pH/sq for the NbN film, the total inductance will be $\sim$1,328 nH. The grounded CPW taper, on the other hand, only has a length of 11 mm (9118 squares) and total inductance of 729 nH. Moreover, the gold ground may serve as a mirror to form an optical cavity with a properly chosen dielectric spacing~\cite{Rosfjord2006}. 

\section{SPICE simulation}
\begin{figure}[H]
    \centering
    \includegraphics[width=.8\linewidth]{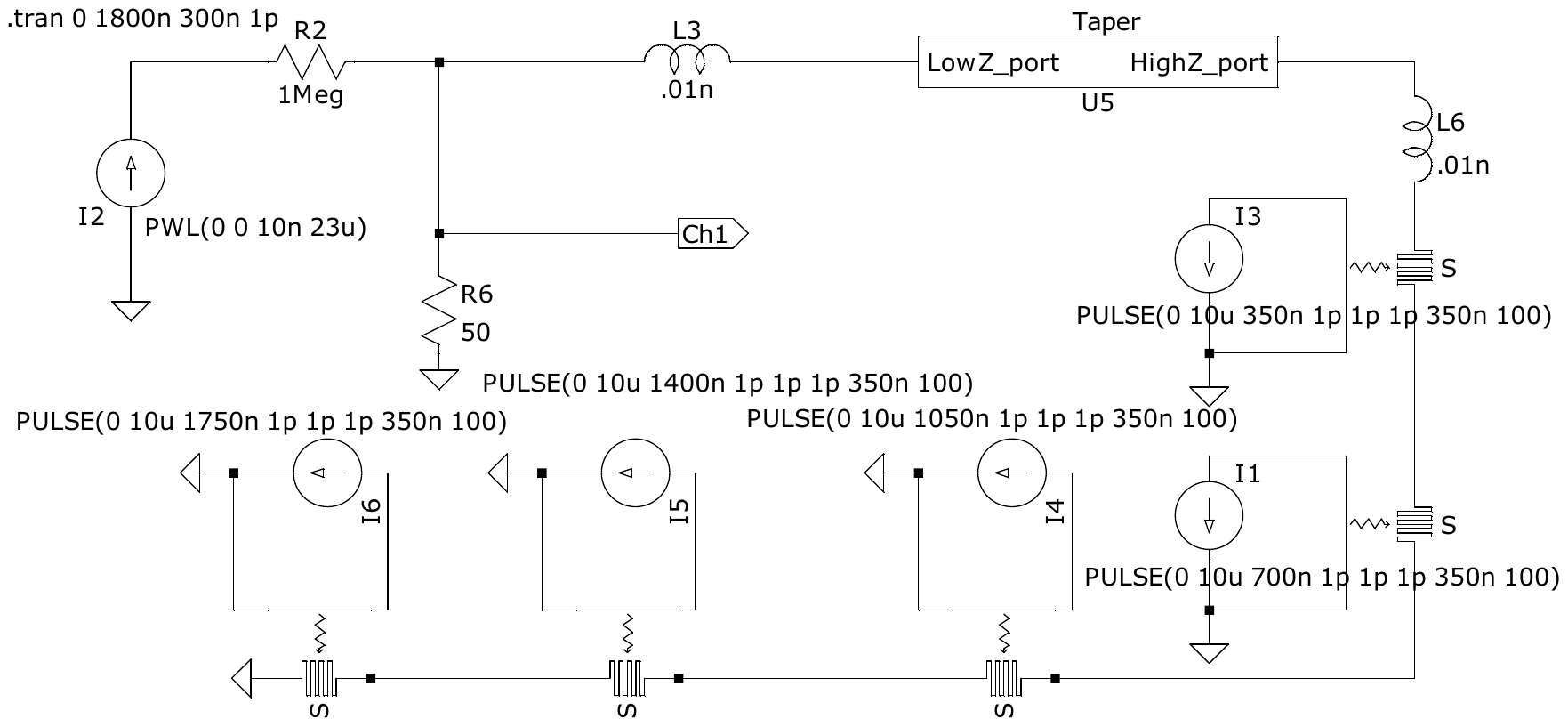}
    \caption{SPICE simulation schematics for multi-photon events in the STaND. The taper is modeled as cascaded transmission lines (300 sections) with varying impedance and phase velocities. To simulate multi-photon events, we divide the nanowire meander into 5 SNSPDs (each with 1/5 inductance) and switch $n$ of them simultaneously. }
    \label{fig:SPICE_schematics}
\end{figure}

We simulate the STaND using a SPICE model that incorporates both the electrothermal feedback and microwave dynamics~\cite{Zhao2018a, Zhu2019, Berggren2018}. To simulate the multi-photon response, we model the nanowire meander as 5 lumped SNSPDs (each with 1/5 of the total inductance) and trigger $n$ of them simultaneously to mimic an $n$-photon event. Figure~\ref{fig:SPICE_schematics} shows the simulation setup. The taper is formed by 300 cascaded transmission lines, each section with impedance and phase velocity set to match the actual taper profile. The SPICE model of each SNSPD is implemented by Berggren et al.~\cite{Berggren2018}, based on the phenomenological hotspot velocity model by Kerman et al.~\cite{Kerman2009}

\begin{figure}[H]
    \centering
    \includegraphics[width = 6 in]{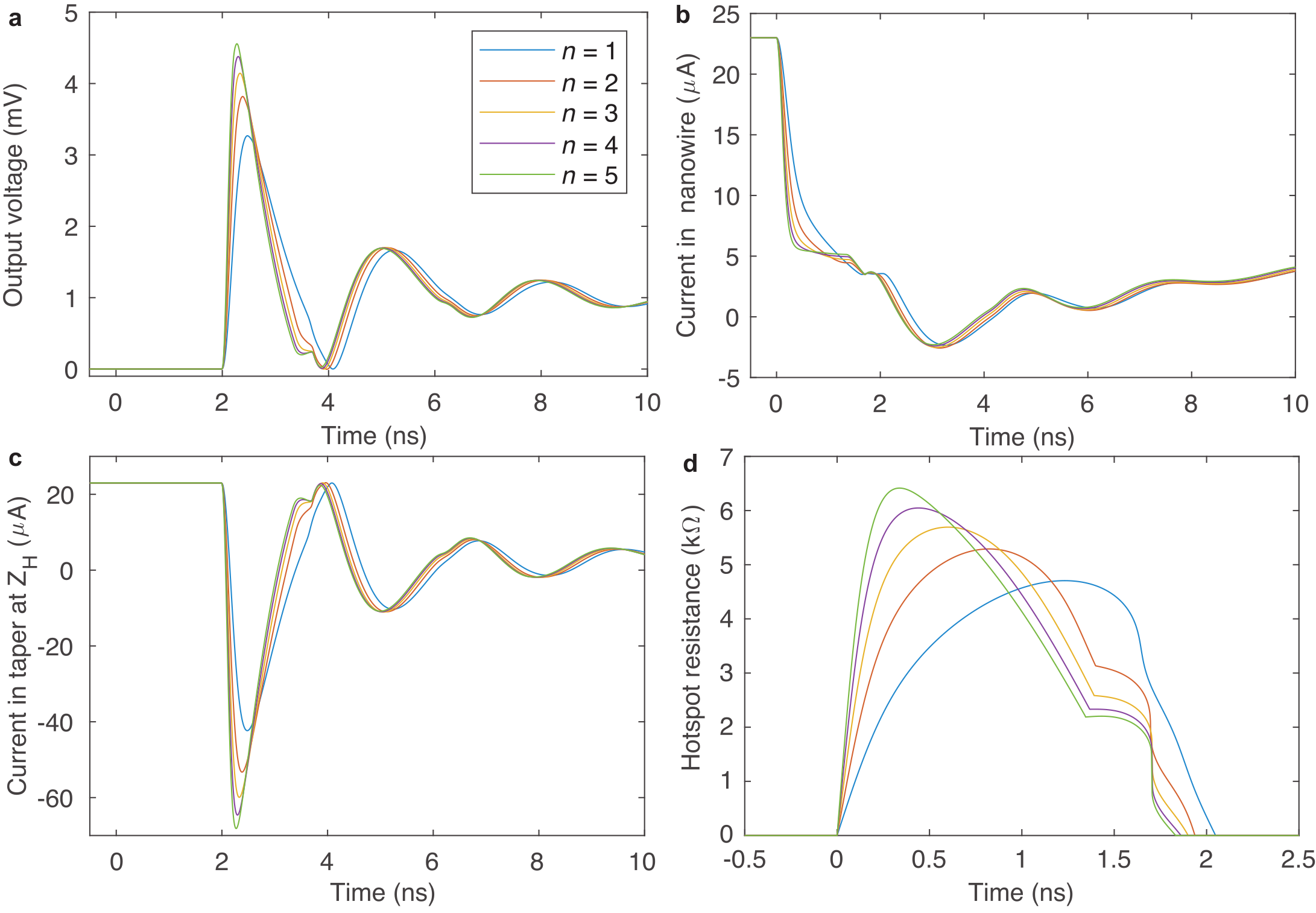}
    \caption{SPICE simulated pulse shapes, current distributions, and hotspot resistances in the STaND. (a) Output voltage on the 50 $\Omega$ load resistor (output voltage); (b) Current in the nanowire, which is also the current at the high-impedance end of the taper (current flowing rightwards are defined as positive); (c) current in the taper at the low impedance (50 $\Omega$) end; (d) evolution of the total hotspot resistance. In general, more photons (i.e., more initial hotspots) create output pulses with larger amplitudes and faster slew rates. This result is qualitatively consistent with our experimental observation. The hotspot resistance increases as $n$, but scales sub-linearly.}
    \label{fig:SPICE_simulation_result}
\end{figure}

Figure~\ref{fig:SPICE_simulation_result} shows the simulation results. Photons arrive at $t=0$ ns. The hotspots start to grow immediately and push current in the nanowire meander towards the taper. After $\sim$2 ns, the current leaves the taper at the low impedance end and enters the 50 $\Omega$ load (and the voltage across the 50 $\Omega$ load is what we see as output). In general, when more photons hit the nanowire, the total hotspot resistance grow faster and larger, and so does the output voltage. However, the maximum hotspot resistance scales sub-linearly as $n$. The taper bandwidth, nanowire inductance, and hotspot growth rate together determine the detector output. 

\section{Supplementary measurement and analysis}

\subsection{Measurement setup}

\begin{figure}[H]
    \centering
    \includegraphics[]{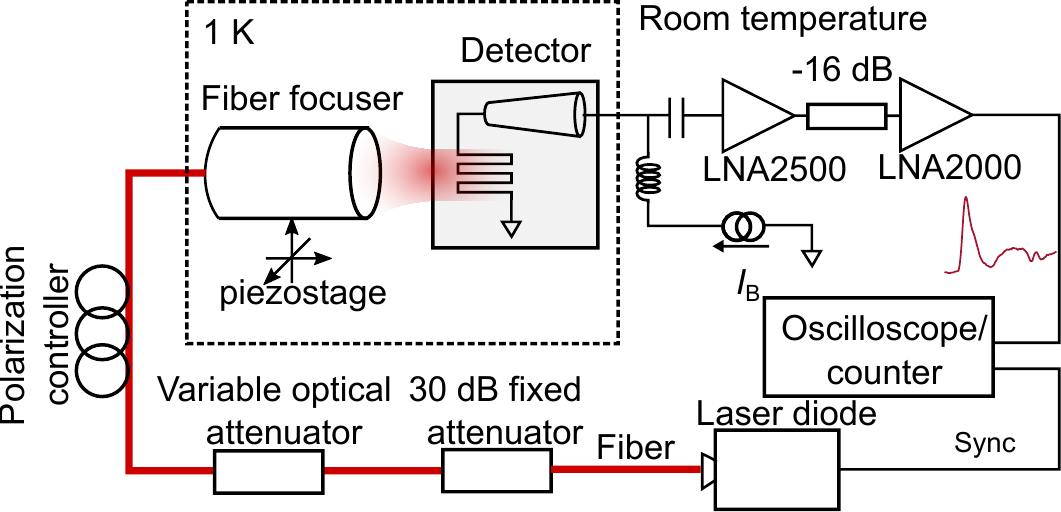}
    \caption{Measurement setup for characterizing the STaND with classical light sources. The fiber-coupled pulsed laser diode (1,550 nm) was attenuated and coupled to the detector chip with a fiber focuser. The fiber focuser was mounted on a piezo-positioner (Attocube) and can move between the reference SNSPD and the STaND. The detectors were read out only using room temperature amplifiers. }
    \label{fig:pnr_setup}
\end{figure}

Figure~\ref{fig:pnr_setup} shows the measurement setup for characterizing the PNR capability of the STaND. A fiber-coupled pulsed diode laser was attenuated (30 dB fixed fiber attenuator in-line with a 0 -- 100 dB calibrated variable attenuator) and coupled to the detector using a fiber focuser. The input polarization was adjusted to maximize detection efficiency. Throughout the measurement, the detector was biased at 23 $\upmu$A. The detector output was amplified using two cascaded room temperature amplifiers. We found that when the count rate was high (close to MHz), the detector would charge the amplifier and the measured output drifts towards larger amplitude. To avoid this effect, we set the laser repetition rate to 100 kHz. In real detector systems where high count rate is necessary, an in-line cryogenic shunt can be added to eliminate this effect. 

When measuring timing jitter of the detectors, we used a 1550 nm mode-locked sub-ps fiber laser. Since the 16 dB attenuator was not necessary for the reference SNSPD, we removed it when measuring timing jitter for the reference SNSPD to increase its signal-to-noise ratio. The timing jitter measurements were all performed in the single-photon regime. 

\subsection{Electrical noise floor of the measurement system}
\begin{figure}[H]
    \centering
    \includegraphics[width=3.5in]{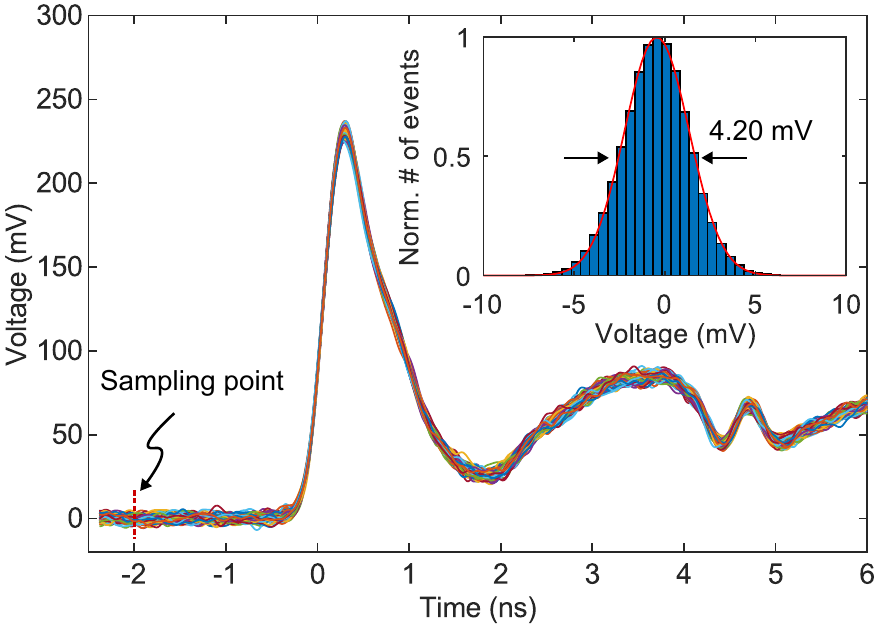}
    \caption{System electrical noise. We sampled the electrical noise on the oscilloscope at 2 ns before the rising edge of the detector pulses. The noise follows a Gaussian distribution with a FWHM of 4.20 mV ($\sigma=1.78$ mV). }
    \label{fig:noise_sampling}
\end{figure}

We sampled the system's electrical noise on the oscilloscope and measured a noise floor of 4.2 mV full-width at half-maximum (see Fig.~\ref{fig:noise_sampling}). This value is smaller than the pulse amplitude distribution in Fig. 2(a) and (b) in the main text. The excessive fluctuation in pulse amplitude may be due to the following factors: (1) Variation of the nanowire widths along the wire causes different hotspot sizes, viz., wider wires generally create smaller hotspots and vice versa. (2) Variation in the location of a detection event along the wire causes differences in microwave dynamics, viz., hotspots near the taper experience different r.f. reflections than the ones near the ground. This effect is particularly strong when the wire length is long, where the lumped element picture breaks down and distributed model becomes appropriate~\cite{Zhao2018a, Zhu2018, Zhao2017}. (3) Variation in inter-arrival time of the photons causes different hotspot evolution, viz., if the second photon is delayed relative to the first photon, it will see a reduced bias current and the hotpot growth will be slower. This effect is described in more detail in Section E below. Furthermore, even if the two photons arrive simultaneously, they may still experience different latencies depending on where they were absorbed across the width of the wire.

\subsection{Rise time and rising-edge slope}

\begin{figure}[H]
    \centering
    \includegraphics[width = 3.5 in]{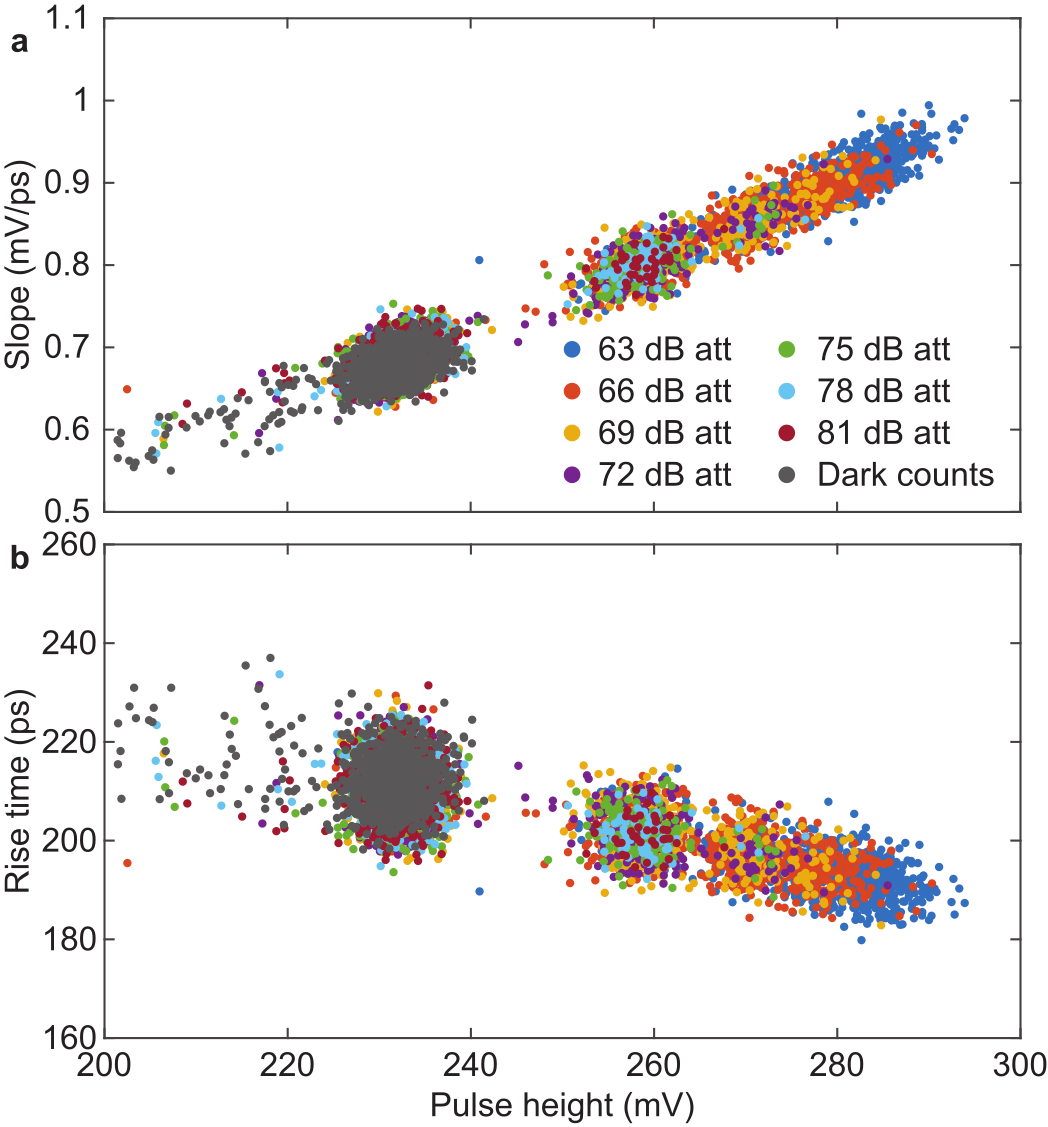}
    \caption{Correlation among pulse height, rising slope (a), and rise time (b). In general, more photons generate pulses with larger amplitude, faster slew rate, and slightly shorter rise time. Pulse amplitude shows the clearest distinction among the three. For each optical attenuation, 1000 pulse traces were recorded and post-processed. The slope was linearly fitted from 40\% to 60\% pulse amplitude, and the rise time was extracted as time take to grow from 20\% to 80\% pulse amplitude. The effective mean photon per pulse at 63 dB attenuation was 5.1.}
    \label{fig:slope_risetime}
\end{figure}

In Fig.~\ref{fig:slope_risetime}, we processed the detector pulses at a range of optical attenuation (63 dB to 81 dB with 3 dB steps; 63 dB corresponded to $\tilde{\mu}=5.1$) and extracted their rising edge slope (slew rate) and rise time. The slope was extracted by linearly fitting the rising edge from 40\% to 60\% pulse amplitude, while the rise time was extracted as 20\% to 80\% time span. 

The slope roughly follows a linear correlation to the pulse amplitude, and thus can also be used to resolve photon numbers, similar to the results reported by Cahall et al.~\cite{Cahall2017} However, in our detector architecture, the slopes are less separable than the amplitudes. As we can see in Fig.~\ref{fig:slope_risetime}(a), the detection events are less separable along the y-axis than along the x-axis. In Fig.~\ref{fig:slope_risetime}(b), the rise time is around 200 ps, and reduces slightly as photon number increases. The changes both in the slope and the rise time qualitatively follow our SPICE simulation---more photons generate detector pulses with both larger amplitude and faster slew rate. 

\subsection{Shoulder in the pulse height histogram}

In the pulse height histograms, we observed a broad shoulder at $<220$ mV. This shoulder is presumed to be from counting events at the nanowire bends, as shown in  Fig.~\ref{fig:bending_crowding}, where the nanowire width gradually increases. In this region, the hotspot cannot grow as large as that in the middle of the meander, and its size has a larger variation due to the range of widths in the bends. Moreover, current tends to crowd in the bends and create ``hot'' corners that are more likely to generate dark counts. To test this hypothesis, we drove the fiber focuser out of focus to illuminate more on the bends and observed increased shoulder that is similar to the dark count histogram, which confirms our hypothesis.

\begin{figure}[H]
    \centering
    \includegraphics[width=3.5in]{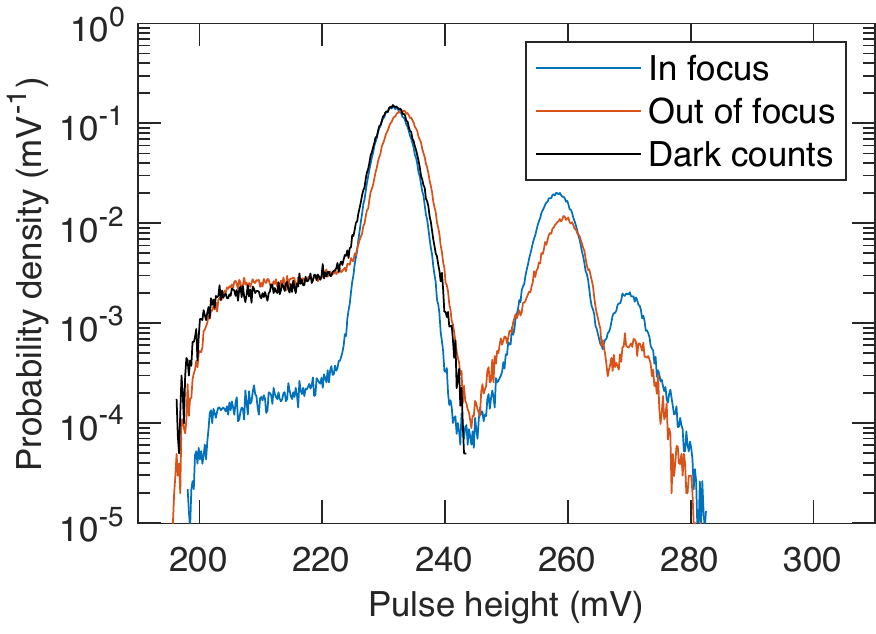}
    \caption{Comparing pulse height distributions under different illumination condition. When we drove the fiber focuser far away from the detector (out of focus), light uniformly illuminated both the wire and bends. In this case, we observed increased shoulder (similar to the dark count case), strongly suggesting that the shoulder originates from the bends.}
    \label{fig:explain_shoulder}
\end{figure}

\begin{figure}[H]
    \centering
    \includegraphics[width = 4.5in]{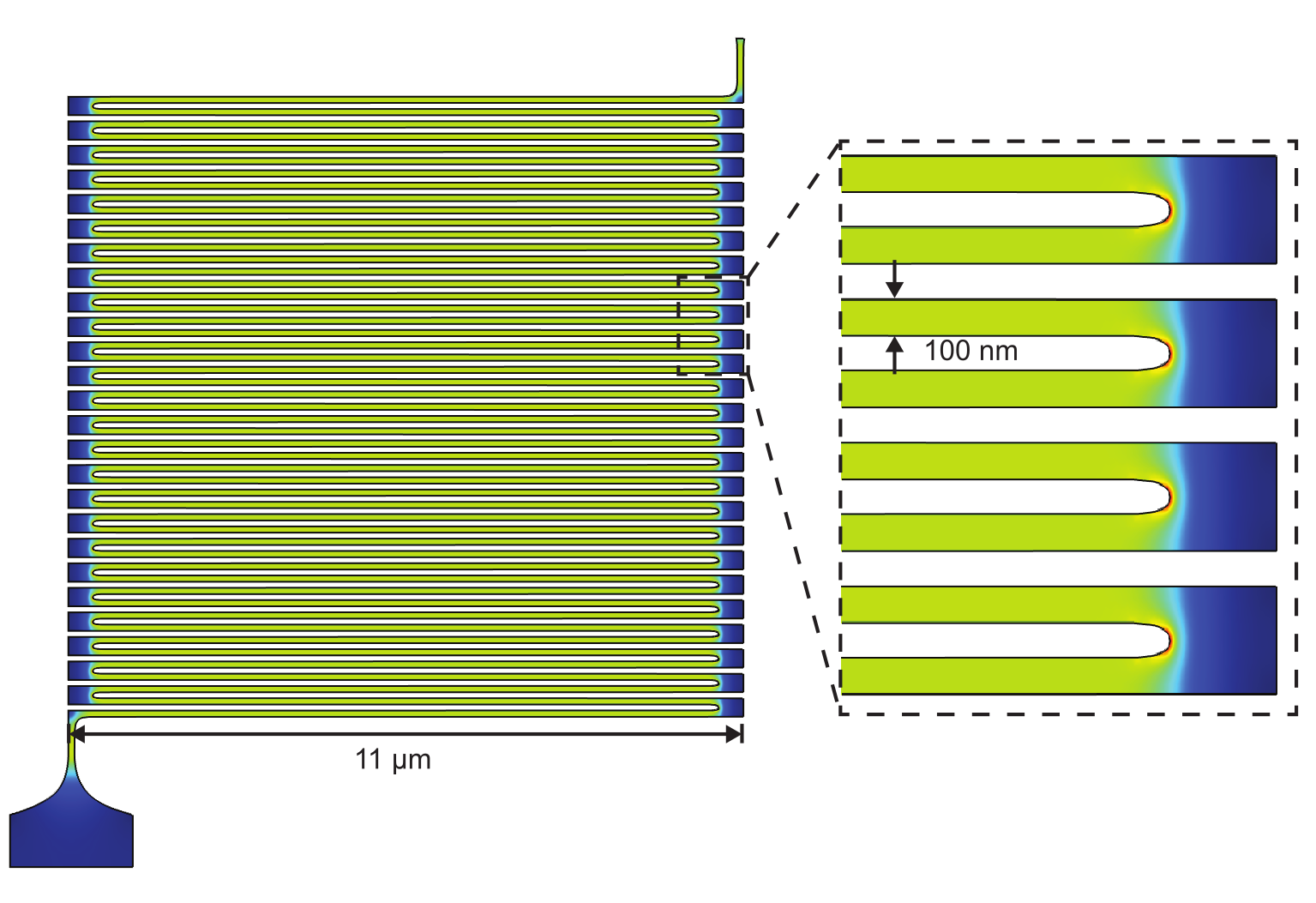}
    \caption{Geometry and current density near the bends. At the bends, the nanowire gradually increases its width. Despite the use of optimized bending curve~\cite{Clem2011}, there are still some current-crowding effects. These areas have higher current density and are prone to generate dark counts. Color represents current density; red: high; blue: low.
    }
    \label{fig:bending_crowding}
\end{figure}

\subsection{Effects of photon inter-arrival time}
The working principle of PNR in the STaND requires multiple photons to arrive close in time, i.e., the photon wavepacket needs to be short. Upon absorption of the first photon, the current in the nanowire starts to drop immediately. It takes about 200 ps for the current to drop to 90\% (inferred from detector rise time in Fig.~\ref{fig:slope_risetime}(b)), and the more initial hotspots, the faster the current drops (Fig.~\ref{fig:SPICE_simulation_result}(b)). If the second photon arrives with some time delay, the nanowire will be at a lower bias current. The second photon will either create a smaller hotspot or fail to initiate a hotspot expansion at all. Therefore, if the laser pulse width is wide (e.g., more than 50 ps), higher-photon events are likely to be underestimated. 

\begin{figure}[H]
    \centering
    \includegraphics[width=3.5in]{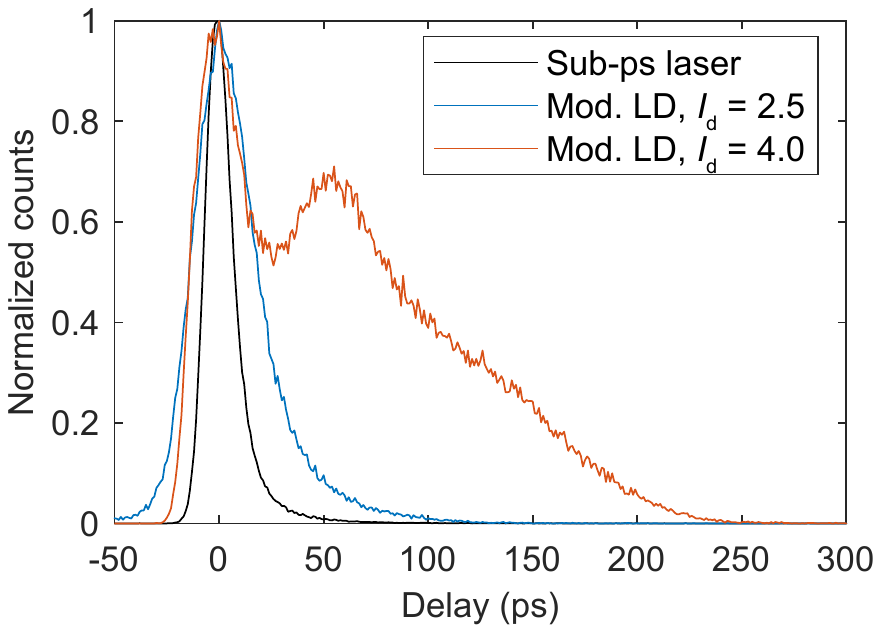}
    \caption{Estimating laser pulse width using the STaND, where the time delay between the laser sync signal and detector pulses were measured. Since the sub-ps laser has negligible pulse width (laser pulse width: 0.18 ps, spectral width: 16.18 nm, fiber dispersion: 18 ps/(nm$\cdot$km), calculated pulse broadening from 2 m fiber: $\sim$0.6 ps), the black curve represents the detector instrument response function. When the modulated laser diode was driven at current setting of 2.5, the pulse width was $\approx33$ ps (FWHM); and when it was driven at 4.0, the pulse broadened significantly ($\approx 100$ ps FWHM, asymmetric with a long tail).}
    \label{fig:laser_width}
\end{figure}

\begin{figure}[H]
    \centering
    \includegraphics{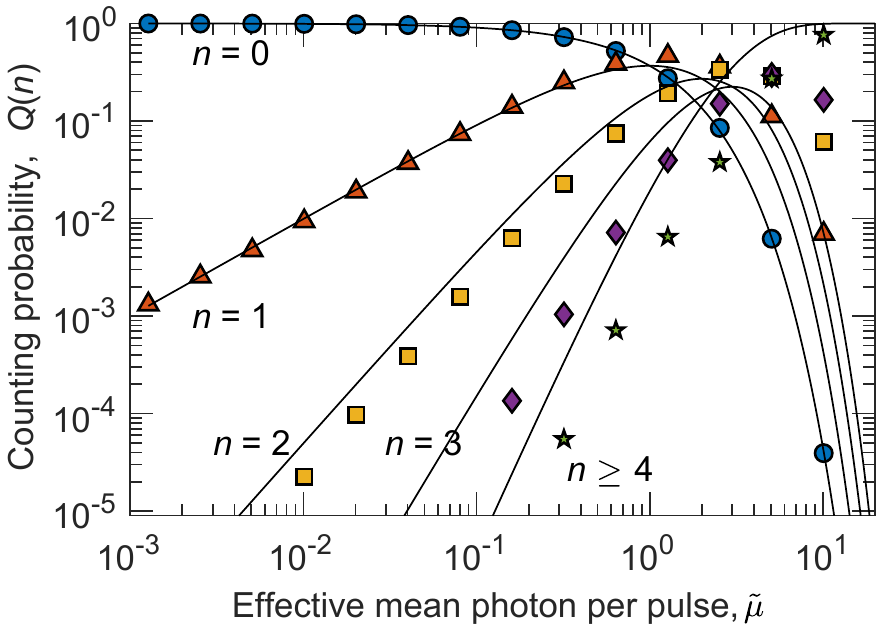}
    \caption{Counting statistics when the modulated laser diode was driven to have a wider ($\approx 100$ ps FWHM, shown as orange curve in Fig.~\ref{fig:laser_width}) pulse width. Higher-photon-number events are significantly under estimated. Symbols: measured data; lines: photon statistics of the source, $S(n) = e^{-\tilde{\mu}}\tilde{\mu}^n/n!$.}
    \label{fig:stats_wide_pulse}
\end{figure}

The pulse width of the modulated laser diode used in our experiments can be changed by tuning the drive current. Since the STaND has a timing jitter as small as 16.1 ps, we used it to estimate the laser pulse width directly. Figure~\ref{fig:laser_width} shows the measured time delay between laser sync signal and detector output when the modulated laser diode was driven at different current settings (these settings are nominal values, and the actual currents were not measured). In the main text, all experiments were performed with the current setting at 2.5, which produced 33 ps wide (FWHM) pulses. When we increase the drive current ($I_\mathrm{d}= 4.0$), the pulse width increased to $\approx100$ ps (FWHM), and the measured photon statistics differed significantly from the expected Poisson statistics of the source (see Fig.~\ref{fig:stats_wide_pulse}), especially for higher photon numbers.

\subsection{Calibrating the comparator readout}
\begin{figure}[H]
    \centering
    \includegraphics[]{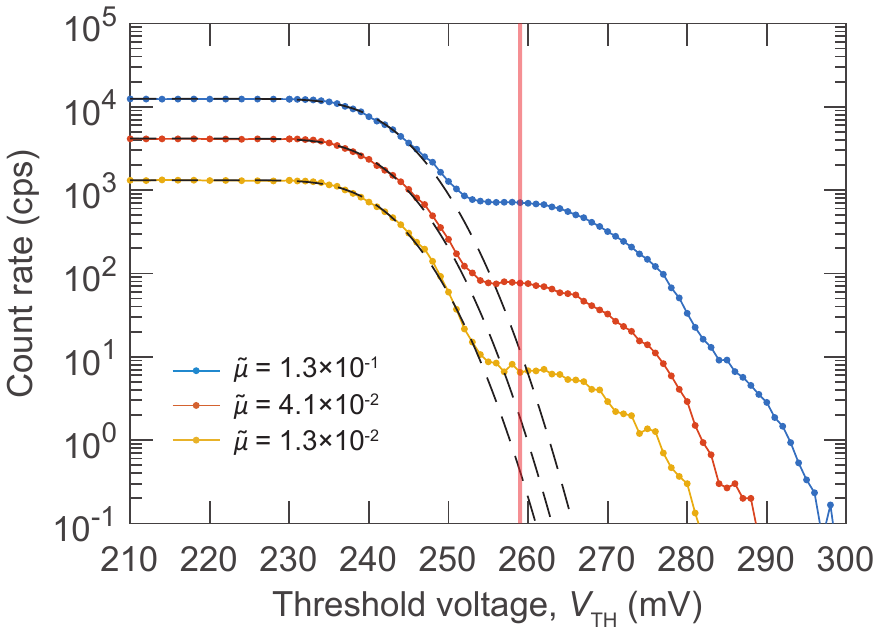}
    \caption{Calibrating the comparator threshold voltage for coincidence counting with a pulsed laser ($f_\mathrm{rep}$ = 100 kHz). Dashed lines are erfc fitting, and the red line at 259 mV marks the chosen $V_\mathrm{TH}$ for coincidence counting.}
    \label{fig:ttl_calibration}
\end{figure}

We calibrated the comparator readout used in Fig. 3 in the main text to find the best threshold voltage for coincidence counting. We illuminated the detector using a pulsed laser with a repetition rate of $f_\mathrm{rep}=100$ kHz. Figure~\ref{fig:ttl_calibration} shows the count rate registered at the counter as a function of $V_\mathrm{TH}$. The dashed lines are complementary error function (erfc) fittings of the roll-offs. The red line marks the chosen threshold voltage $V_\mathrm{TH}=259$ mV for the coincidence counting. From the erfc fitting, we found $V_\mathrm{TH} = 259$ mV was 3.28$\sigma$ away from the single-photon main peak (241 mV).  These values differed from the ones measured using the oscilloscopes (e.g., in Fig.1 and Fig.2 in the main text) because the comparator had limited bandwidth (300 MHz)  and distorted the pulse shapes.  The limited bandwidth and noise performance of the comparator had also degraded the level discrimination integrity. Nevertheless, the comparator readout was faster than post-processing and had finer threshold resolution than the counter's internal trigger settings. 

\section{Basic detector metrics}
\subsection{Efficiency and dark count rate}
\begin{figure}[H]
    \centering
    \includegraphics[]{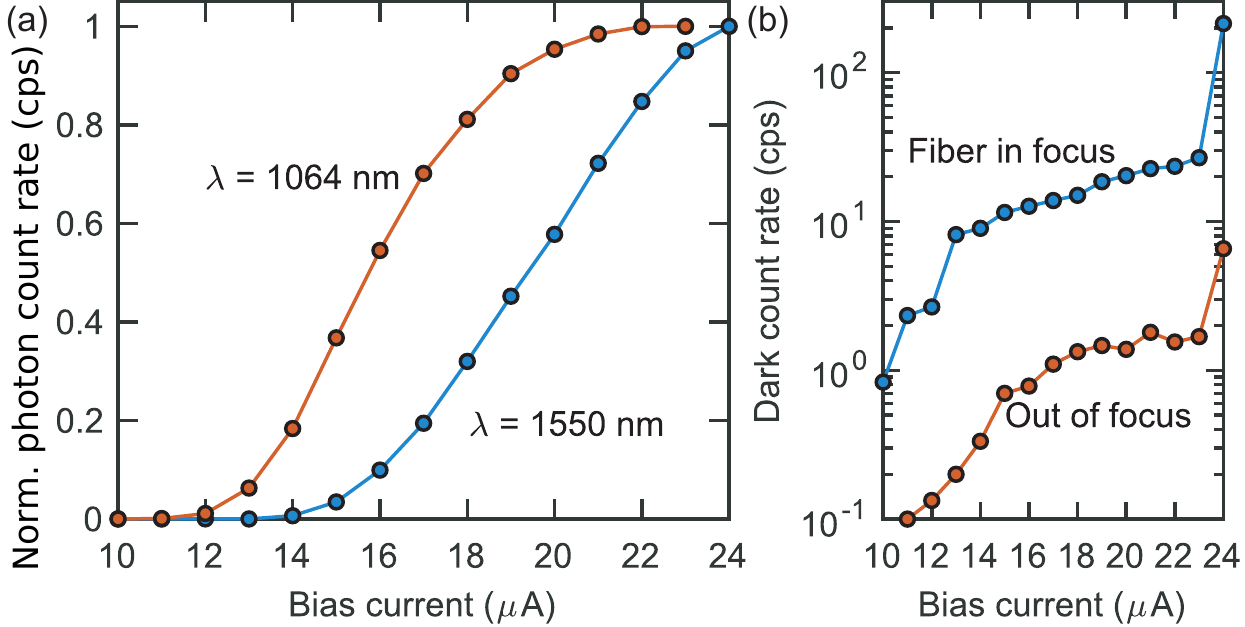}
    \caption{Normalized Photon count rate (PCR) and dark count rate (DCR) as functions of bias current.}
    \label{fig:pcr_dcr}
\end{figure}

Figure~\ref{fig:pcr_dcr}(a) shows the normalized photon count rate (PCR) as a function of bias current. Under 1064 nm illumination, the detector showed saturated quantum efficiency; and at 1550 nm, it passed its inflection point and was close to saturation. At 23 $\upmu$A with 1550 nm illumination, the STaND ($11\,\upmu\mathrm
{m}\times10\,\upmu\mathrm{m}$ area with 50\% fill-factor) had a system detection efficiency (including coupling loss up to the fiber feedthrough at the cryostat) of $\approx 5.6\%$.

Figure~\ref{fig:pcr_dcr}(b) shows the dark count rate (DCR). When the fiber focuser was moved away, the DCR dropped by one order of magnitude, indicating that the DCR was dominated by leakage photons channeled through the fiber. At $I_\mathrm{B} = 23 \upmu$A, the system dark count rate (fiber in focus) was 26.8 c.p.s., and the device dark count rate (fiber out of focus) as 1.7 c.p.s.

\subsection{Reset time}

We estimate the detector's reset time from the pulse decay. The reset time of the SNSPDs and STaNDs are limited by the kinetic inductance, and the output pulse follows an exponential decay $\exp(-t/\tau)$, where $\tau = L/R$. $L$ is the total inductance of the device, including both the nanowire meander and the taper, and $R = 50\,\Omega$ is the load impedance of the readout circuitry. 

\begin{figure}[H]
    \centering
    \includegraphics[]{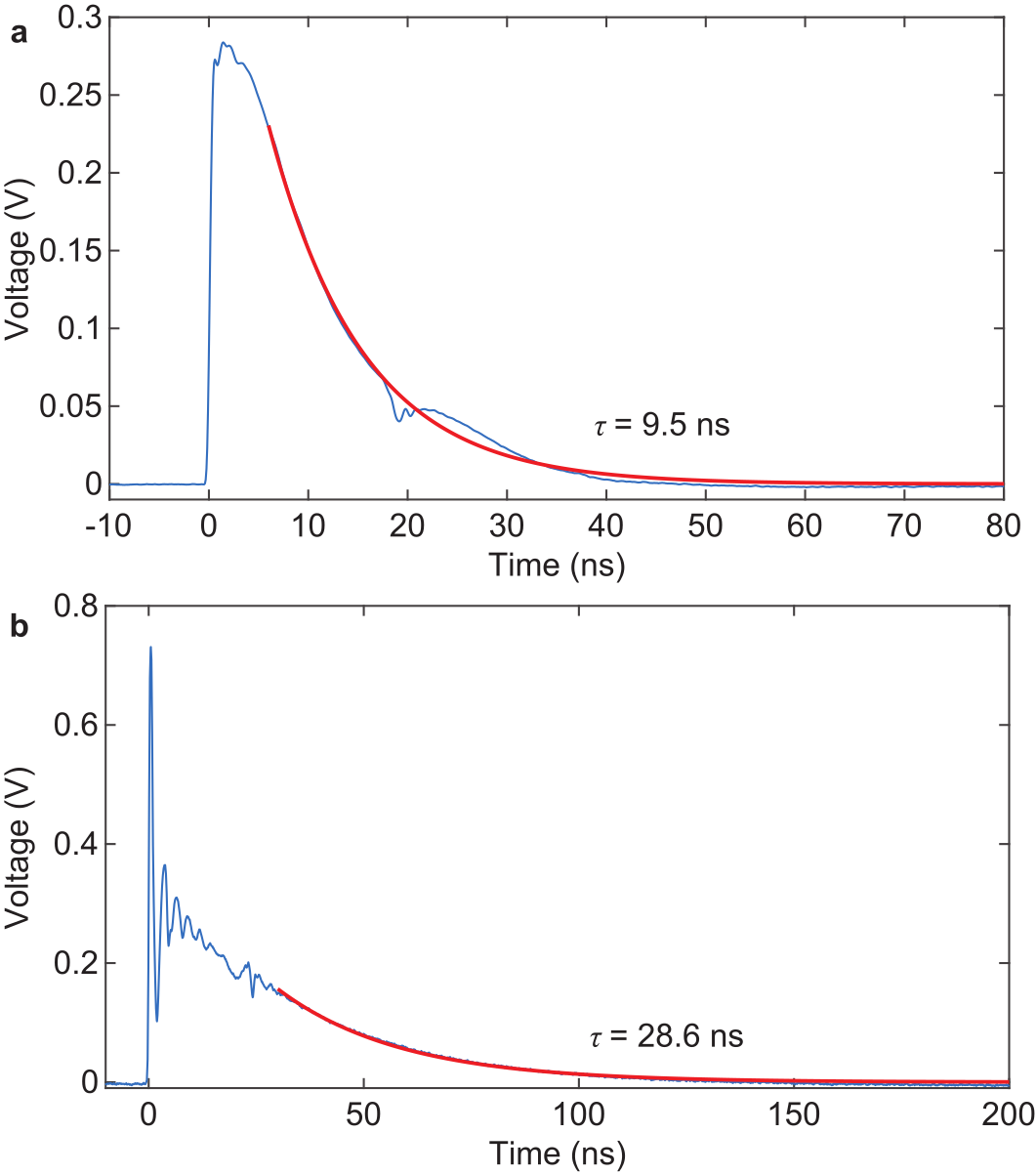}
    \caption{Estimating reset time from pulse decay. Exponential fitting ($e^{-t/\tau}$, where $\tau = L/R$) of pulse decay shows that reference SNSPD has $\tau$ of 9.5 ns (a), and the STaND has $\tau$ of 28.6 ns.}
    \label{fig:reset_time}
\end{figure}

Figure~\ref{fig:reset_time} shows the averaged pulse shapes of the SNSPD (a) and STaND (b). These pulses were amplified using a low-frequency amplifier (MITEQ AM-1309, gain: 50 dB, bandwidth: 1 kHz - 1 GHz) instead of the ones used in the main text because the 1 kHz lower cut-off would ensure accurate capture of the slow decay process. Note that this amplifier was saturated and lost some high-frequency features on the rising edge, but this saturation did not affect our analysis on the falling tail. Exponential fitting of the falling tails gave the $L/R$ time constants for the SNSPD and STaND to be 9.5 ns and 28.6 ns, respectively. With $1/(3\tau)$ as a rule of thumb, their maximum count rates were 35.1 MHz and 11.7 MHz, respectively.

The SNSPD was designed to be 5,200 squares, and the STaND was designed to be 21,800 squares (i.e., the taper was 16,600 squares). The fitted $L/R$ time constants did not strictly follow the ratio of the device's number of squares. This may be due to (1)  the nanowire meander had larger sheet inductance due to the presence of near-switching bias current, or (2) fabrication error that led to discrepancy in device geometry.

\section{Counting statistics and estimation of effective mean photon (per pulse)}

Here we show, in the case of coherent state illumination, that the coupling loss and detector efficiency can be treated as an effective attenuation to the source, and the effective mean photon $\tilde{\mu} = \eta\mu$ can be estimated by fitting the photon count rate as a function of the known variable optical attenuation applied to the pulsed laser source. 

A uniformly illuminated STaND can be treated as a spatially-multiplexed, $N$-element ($N$ is on the order of 1,000), uniform detector array. Such a detector array is usually modeled as an $N$-port beam splitter, where each output port is coupled to a single-photon detector with efficiency $\eta$. For $n$-photon input, the probability of no-click is $P_\eta^N(0|n) = (1-\eta)^n$, and the probability of correctly getting the photon number is $P_\eta^N(n| n)=\left(\frac{\eta}{N}\right)^{n} \frac{N !}{(N-n) !}$, for $n \leq N$. The cases in between, i.e., $n$ photon input but detector tells $k$, can be solved recursively~\cite{Fitch2003,Dauler2009b},
\begin{equation}
P_{\eta}^{N}(k | n)=\binom{N}{k} \sum_{j=0}^{k}(-1)^{j}\binom{k}{j}\left[(1-\eta)+\frac{(k-j) \eta}{N}\right]^{n}
\end{equation}
where $\binom{N}{k}=N!/[k! (N-k)!]$.   

For coherent source illumination with a mean photon number of $\mu$, the counting probability follows
\begin{equation}
\begin{aligned} Q(k) &=\sum_{n=0}^\infty P_\eta^N(k|n) S_\mu(n)\\&=\sum_{n=0}^{\infty}\binom{N}{k} \sum_{j=0}^{k}(-1)^{j}\binom{k}{j}\left[(1-\eta)+\frac{(k-j) \eta}{N}\right]^{n} \frac{e^{-\mu}\mu^n}{n!}\\ &=\binom{N}{k} e^{\frac{\eta  \mu  (k-N)}{N}} \left(1-e^{-\frac{\eta  \mu }{N}}\right)^k, 
\end{aligned}
\end{equation}
where $S_\mu(n) =  e^{-\mu}\mu^n/n!$ is the Poissonian photon statistics of a coherent source.

Now, if we illuminate a unit-efficiency detector array using coherent source with mean photon $\tilde{\mu} = \eta\mu$, the counting probability will be
\begin{equation}
\begin{aligned}
Q'(k) & = \sum_{n = 0}^\infty P_{\eta = 1}^N(k|n)S_{\tilde{\mu}}(n) \\&= \sum_{n=0}^{\infty}\binom{N}{k} \sum_{j=0}^{k}(-1)^{j}\binom{k}{j}\left[\frac{(k-j)}{N}\right]^{n} e^{-\tilde{\mu}} \frac{\tilde{\mu}^{n}}{n !} \\&=\binom{N}{k} e^{\frac{\eta  \mu  (k-N)}{N}} \left(1-e^{-\frac{\eta  \mu }{N}}\right)^k, 
\end{aligned}
\end{equation}
which is identical to $Q(k)$, meaning that the counting statistics is equivalent between the two cases. Note that when $N\gg k$, $Q(k) \approx  e^{\tilde{\mu}}\tilde{\mu}^k/k!$, which is appropriate for Fig. 2(c) in the main text.

\begin{figure}[H]
    \centering
    \includegraphics[]{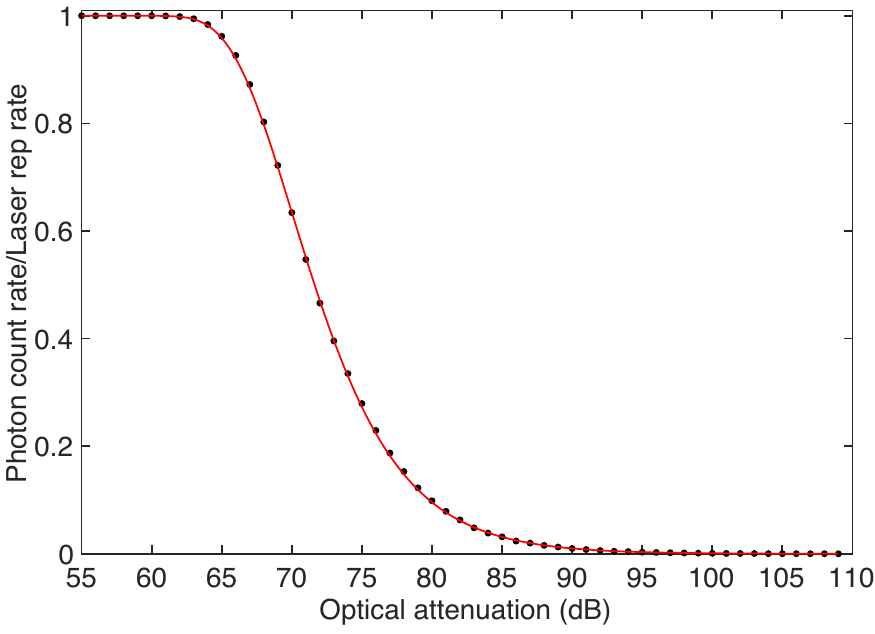}
    \caption{Estimating effective mean photon per pulse $\tilde{\mu}$. The counter threshold was set below the single-photon pulse amplitude to capture all $k\ge 1$ events. For coherent state illumination, the counting/clicking probability $P_\mathrm{click} = PCR/f_\mathrm{rep}=1-\exp(-\gamma\tilde{\mu})$, where $PCR$ is the photon count rate, $f_\mathrm{rep}$ is the laser repetition rate. By fitting the counting probability as a function of applied external optical attenuation, we get $\tilde{\mu}=10.080\pm0.050$ at 60 dB attenuation (uncertainty indicates 95\% confidence bound).}
    \label{fig:mean_photon_rate}
\end{figure}

To estimate $\tilde{\mu}$ experimentally, we set the trigger level of the counter below the single-photon pulse amplitude and measured the photon count rate ($PCR$) as a function of applied optical attenuation ($\gamma$), as shown in Fig.~\ref{fig:mean_photon_rate}. Since the clicking probability $PCR/f_\mathrm{rep}$ is essentially $Q(k\ge 1) = 1- Q(k=0) $, we fit it with $1-\exp(-\gamma\tilde{\mu})$ and get $\tilde{\mu}=10.080\pm0.050$ at 60 dB attenuation. The accuracy of this method was ensured by the stability of laser power and repetition rate as well as calibration of the variable optical attenuator. The use of $\tilde{\mu}$ in analyzing the measured counting statistics helped us isolate the detector's intrinsic architectural limit on PNR from external factors that could be later optimized, such as optical coupling loss and absorption efficiency.
 
\section{STaND array vs. SNSPD array for photon number resolution}

\begin{figure}[tbh!]
    \centering
    \includegraphics[width=\textwidth]{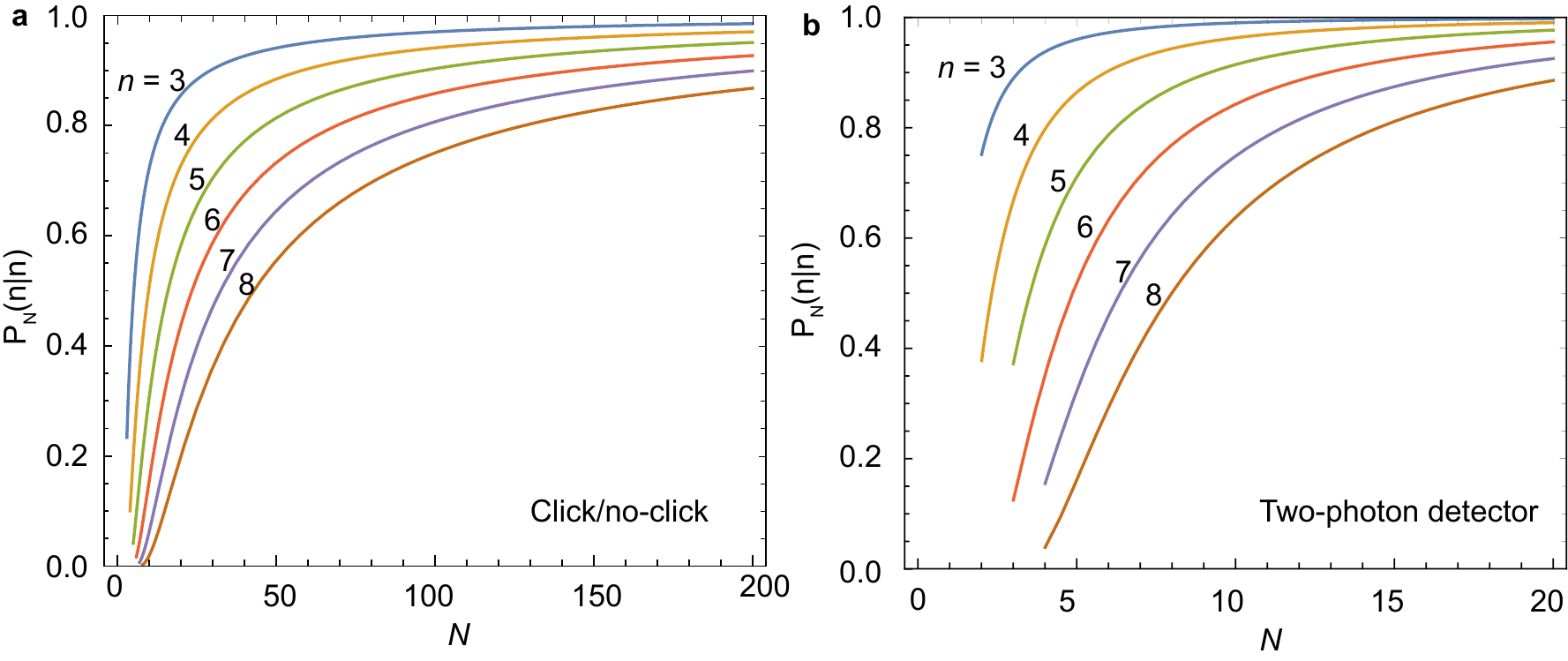}
    \caption{Comparison of resolving fidelity of large photon numbers using arrays of SNSPDs and STaNDs. Here, we assume unity detection efficiency. To include actual efficiency, a scaling factor of $\eta^n$ needs to be multiplied to both cases.}
    \label{fig:arrayprobability}
\end{figure}

Similar to spatially multiplexed SNSPD arrays, it is possible to use arrays of STaNDs to resolve larger number of photons. Here we treat the STaND as a perfect two-photon detector, and compare the PNR fidelity of $N$-element STaND arrays against $N$-element click/no-click SNSPD arrays. The probability of correctly resolving an $n$-photon input in an $N$-element SNSPD array follows $P_N^{\mathrm{SNSPD}}(n|n) = \eta^n N!/[N^n(N-n)!]$ for $N\ge n$, i.e., no two-or-more photons hit the same element. For an $N$-element STaND array, we demand no three-or-more photons hit the same element. For instance, neglecting the $\eta^n$ term for all cases, $P_N^\mathrm{STaND}(3|3) = 1 - 1/N^2$, $P_N^\mathrm{STaND}(4|4) = 1 - \frac{1+4(N-1)}{N^3}$, $P_N^\mathrm{STaND}(5|5) = 1 - \frac{1+5(N-1)+10(N-1)^2}{N^4}$, and so on ($N\ge \mathrm{ceil}({n/2})$). We plot them in Fig.~\ref{fig:arrayprobability} assuming $\eta = 1$. On average, to achieve similar fidelity (e.g., 90\% with $\eta=1$), one needs roughly 10 times more SNSPDs than STaNDs.

